\documentclass{aa}
\usepackage{epsfig}

\usepackage{amsmath}
\usepackage{amssymb}
\usepackage{graphics}
\usepackage{amstext}
\usepackage{txfonts}
\usepackage{graphicx}

\begin{document}

\title{On type Ia supernovae and the formation of \\ single low-mass
white dwarfs}
\author{Stephen Justham\inst{1}\thanks{Now at: The Kavli Institute
    for Astronomy and Astrophysics, Peking University, Beijing, China
    {\tt (sjustham@kiaa.pku.edu.cn)}} \and Christian Wolf\inst{1}
  \and Philipp Podsiadlowski\inst{1} \and Zhanwen Han\inst{2}}

\institute{Department of Physics, Oxford University, Keble Road,
Oxford, OX1 3RH, United Kingdom {\tt (sjustham@astro.ox.ac.uk)}
\and National Astronomical Observatories / Yunnan
Observatory, Chinese Academy of Sciences, Kunming, 650011, China}

\date{Received .... Accepted ....}

\abstract{There is still considerable debate over the progenitors of
type Ia supernovae (SNe Ia). Likewise, it is not agreed how single
white dwarfs with masses $\rm \lesssim 0.5 ~M_{\odot}$ can be formed
in the field, even though they are known to exist.}{We consider
whether single low-mass white dwarfs (LMWDs) could have been formed in
binary systems where their companions have exploded as a SN Ia. In
this model, the observed single LMWDs are the remnants of giant-branch
donor stars whose envelopes have been stripped off by the supernova
explosion.}{We investigate the likely remnants of SNe Ia, including
the effects of the explosion on the envelope of the donor star.  We
also use evolutionary arguments to examine alternative formation
channels for single LMWDs. In addition, we calculate the expected
kinematics of the potential remnants of SNe Ia.}{SN Ia in systems with
giant-branch donor stars can naturally explain the production of
single LMWDs. It seems difficult for any other formation mechanism to
account for the observations, especially for those single LMWDs with
masses $\rm \lesssim 0.4 ~M_{\odot}$. Independent of those results, we
find that the kinematics of one potentially useful population
containing single LMWDs is consistent with our model. Studying remnant
white-dwarf kinematics seems to be a promising way to investigate SN
Ia progenitors.}{The existence of single LMWDs appears to constitute
evidence for the production of SNe Ia in binary systems with a
red-giant donor star. Other single white dwarfs with higher space
velocities support a second, probably dominant, population of SN Ia
progenitors which contained main-sequence or subgiant donor stars at
the time of explosion. The runaway stars LP400-22 and US 708 suggest
the possibility of a third formation channnel for some SNe Ia in
systems where the donor stars are hot subdwarfs.}


\maketitle

\titlerunning{Single LMWDs as SN Ia remnants}
\authorrunning{S.~Justham et al.}

\section[]{Introduction}

Type Ia supernovae (SNe Ia) are of major astrophysical importance.
They have acquired particular cosmological significance since they
have been used to measure the expansion history of the Universe (Riess
et al.\ 1998; Perlmutter et al.\ 1999; Riess et al.\
2004). Understanding their nature is also of importance for
understanding the metallicity evolution and star-formation history of
galaxies (e.g. Canal, Ruiz-Lapuente \& Burkert 1996; Matteucci \&
Recchi 2001). Despite their importance, there is still no agreement on
the nature of their progenitors.

There is broad agreement that the destruction of a white dwarf (WD) in
a thermonuclear explosion constitutes the supernova event itself, but
there are two main classes of competing models for the events which
lead to the explosion. In the single-degenerate scenario, the doomed
WD accretes matter from a non-degenerate companion (Whelan \& Iben
1973; Nomoto 1982; Han \& Podsiadlowski 2004). In the
double-degenerate scenario, the mass donor is a second WD; the most
commonly considered scenario involves the merger of two CO WDs (Iben
\& Tutukov 1984; Webbink 1984; also see Martin, Tout \& Lesaffre 2006
for a variant of this scenario). An explosion following the merger of
two WDs would leave no remnant, whilst the companion star in the
single-degenerate scenario would survive and be potentially
identifiable (Ruiz-Lapuente 1997; Podsiadlowski 2003; Ruiz-Lapuente et
al.\ 2004).

There has been no conclusive proof to date that any individual object
is the surviving non-degenerate donor from a SN Ia explosion. If
Chandrasekhar mass WD-WD mergers do {\emph{not}} lead to SNe Ia, they
are expected to leave a remnant neutron star via acretion-induced
collapse (AIC, see Nomoto \& Iben 1985; Nomoto \& Kondo 1991; but see
also Yoon, Podsiadlowski \& Rosswog 2007). At present we do not know
whether {\emph{all}} WD--WD mergers {\emph{do}} leave remnants -- in
which case the double degenerate scenario could not be responsible for
SNe Ia -- and it seems unlikely that that this will become clear in the
near future (but see, e.g., Levan et al.\ 2006).  Hansen (2003) first
noticed that observed high-velocity WDs (Oppenheimer et al.\ 2001)
could have been produced through SNe Ia; such WDs would be the
descendants of non-degenerate mass donors in the pre-supernova
binaries.  Hansen's idea seems to be consistent with more detailed
work on the ages of the WDs in the Oppenheimer et al.\ sample by
Bergeron et al.\ (2005) and deserves further attention, but by itself
it is not a clinching argument for the single-degenerate channel. Nor
has the evidence that the SN Ia rate is different for different
stellar populations (Mannucci et al.\ 2005) led to firm
conclusions. The strongest direct evidence that non-degenerate donor
stars can lead to normal type Ia supernovae has been provided by Patat
et al.\ (2007), who observed circumstellar material around SN 2006X
which seems extremely hard to reconcile with a double-degenerate
progenitor.

Here we suggest that the observed, apparently single, low-mass white
dwarfs (LMWDs) provide evidence that at least some SN Ia explosions
have occurred with non-degenerate donor stars. We define LMWDs as WDs
which are too low in mass to have been produced by single-star
evolution as we currently understand it. A population of single LMWDs
has been implied by, e.g., the work of Maxted, Marsh \& Moran
(2000).\footnote{See also Marsh, Dhillon \& Duck (1995); their study
was originally motivated by their interest in finding double
degenerate binaries as potentional SN Ia progenitors.}  We also
investigate the apparently single ultra-cool white dwarfs (UCWDs) as
potentially containing a useful subset of the LMWD population and
indicate how further observations of the kinematics of this and other
populations could lead to constraints on the progenitors of SNe Ia.

In sections \ref{sec:LMWDchannels} and \ref{sec:SNIaPops} we argue
that the existence of single LMWDs is most naturally explained by the
single-degenerate model for SNe Ia. In section \ref{sec:UCWDs} we
introduce UCWDs, discuss to what extent the observed single UCWDs
might be a useful sample of LMWDs, and consider in what way the
observed single UCWD population is consistent with a SN Ia origin.

\section[]{Formation channels for single LMWDs}
\label{sec:LMWDchannels}
 
Current evidence suggests that $\rm 1 ~M_{\odot}$ zero-age
main-sequence (ZAMS) stars, left to evolve in isolation, produce white
dwarfs $\rm\ga 0.55 ~M_{\odot}$ (e.g. Han, Podsiadlowski \& Eggleton
1994; Weidemann 2000). In order to produce low-mass helium WDs (with
masses $\rm \lesssim 0.5 ~M_{\odot}$), it is necessary to remove the
envelope of a star before it is able to ignite helium. Within the age
of the Universe, it is almost certainly impossible for a single star
to produce WDs with masses close to $\rm0.3 ~M_{\odot}$, or even $\rm
0.17 ~M_{\odot}$, as recently inferred for the runaway WD LP 400-22
(Kawka et al.\ 2006), or $\rm 0.23 ~M_{\odot}$, the estimated mass of
the apparently single UCWD LHS 3250 (Bergeron \& Legget 2002).


\subsection{Production of single LMWDs by single stars?}

It would be simplistic to conclude from the existence of currently
single LMWDs that single stars can produce LMWDs. However, arguments
have been made in favour of a single star channel for the production
of some LMWDs, at least in significantly metal-rich populations
(Kalirai et al.\ 2007; Kilic, Stanek \& Pinsonneault 2007; see also
Han et al.\ 1994). Whilst this is a possibility, that interpretation
is not the only one which can explain the observations: this paper
provides one alternative applicable to both field stars and clusters,
and dynamical interactions in dense clusters may provide another route
(e.g. Adams, Davies \& Sills 2004).\footnote{Note that mass
segregation tends to move low-mass objects outwards in such clusters,
so LMWDs might be formed in the core and observed far from the
centre.} Bedin et al.\ (2008) and Van Loon, Boyer \& McDonald (2008)
do not support the single-star formation channel for LMWDs suggested
by Kalirai et al.\ (2007). \footnote{Bedin et al.\ argue that
there may not, after all, be an unusual single LMWD population in NGC
6791 and Van Loon et al.\ see no evidence for enhanced stellar mass
loss in infra-red observations of the cluster.}

Even if some WDs less massive than $\rm\sim 0.55 ~M_{\odot}$ can be
produced by single-star evolution with super-solar metallicity, the
masses are unlikely to approach those of LP 400-22 or LHS 3250 (as
above). Follow-up calculations to Han et al.\ (1994) imply that only
LMWDs with $ M\gtrsim ~0.4 ~{\rm M_{\odot} }$ might be produced from
such a single-star channel, even at high metallicity (Meng, Chen \&
Han, 2007). Kilic et al.\ (2007b) also state that their proposed
single star channel is highly unlikely to be relevant for WD masses as
low as $\rm 0.2 ~M_{\odot}$.

Nothing of what follows would be significantly affected if the maximum
mass of the LMWD category had to be revised downwards slightly to
account for non-standard single-star evolution.  We will mostly be
considering LMWDs with masses $\rm \lesssim 0.3 ~M_{\odot}$, at which
point any presently proposed single-star channel is not expected to
contribute.

\subsection{Production of single LMWDs by binary stars?}

In binary systems we can invoke mass transfer (and sometimes ablation
by a pulsar companion) in order to explain the observed binary LMWDs
(e.g. van Kerkwijk et al.\ 2000; Liebert at al. 2004). Apparently
\emph{single} LMWDs must also be formed within an interacting binary
system -- either we have not detected their companion or the binary
has been disrupted. One attractive formation channel stands out: the
formation of the LMWD in a binary where the binary companion exploded
in a SN Ia. Before examining this channel in more detail, we discuss
possible alternative explanations for single LMWDs.

\subsubsection{Alternatives to SN Ia: Core-collapse supernovae?}

The natural alternative to a SN Ia in explaining the disruption of a
binary is a core-collapse supernova (chosen such that the system
becomes unbound). However, forming a LMWD in such a system is
challenging. The binary must remain intact long enough for the WD
progenitor to lose its envelope such that it will later form a LMWD;
this implies that the initially more massive star has transferred its
envelope to the secondary and then becomes a WD. In order to form a
LMWD, the primary mass must be $\rm \leqslant 4.5 ~M_{\odot}$, which
our own stellar calculations find to have a core mass at the end of
the main sequence of $\rm \approx 0.52~M_{\odot}$.\footnote{We use
Eggleton's stellar evolution code (Eggleton 1971; Pols et al.\ 1995)
with a metallicity of 0.02 along with the convective overshooting
calibration of Pols et al.\ (1998).} In order to produce a
core-collapse supernova (requiring $ M_{\rm ZAMS}~\gtrsim~8~{\rm
M_{\odot}}$), the initially less massive star would thus have to
accrete the large majority of the envelope of the primary soon after
the primary has left the main sequence, and the initial mass ratio
would have to be close to 1. This highly optimistic scenario does not
produce a distinctly low-mass WD; to produce a $\rm 0.3~M_{\odot}$ WD
in this way requires a ZAMS mass for the primary of $\rm \approx
3~M_{\odot}$, precluding a core-collapse supernova in the binary
system.

Our arguments above are generous; Davies, Ritter \& King (2002) found
a much more restrictive result. In a different context they
investigated the evolution of systems where a WD is formed before a
core-collapse supernova occurs in the system. They concluded that
`... the mass of the white dwarfs generated in this way, $ M_{\rm WD}
\gtrsim 1~{\rm M_{\odot}}$'.

\subsubsection{Alternatives to SN Ia: Circumbinary discs?}

If cataclysmic-variable (CV) evolution is driven by circumbinary
discs, the donor star may eventually be entirely consumed (the `White
Widow' scenario; see Spruit \& Taam 2003, following Spruit \& Taam
2001 and Taam \& Spruit 2001). For this mechanism to explain single
LMWDs, the WD in the progenitor CV must also have been a LMWD; the WD
may not gain much mass, as the matter it accretes can be ejected via
nova explosions, but it is unlikely to become significantly less
massive. 

It is not clear whether this mechanism operates in CVs: attempts to
detect circumbinary discs have inferred disc masses several orders of
magnitude below the required values (Dubus et al.\ 2004; Muno \&
Mauerhan 2006). Hence in the abscence of further supporting evidence
we consider this potential formation channel unlikely at present.

\subsubsection{Alternatives to SN Ia: Accretion-induced collapse?}

Some systems will contain an accreting WD which succeeds in reaching
the Chandrasekhar mass but fails to produce a supernova as the WD is
predominantly composed of oxygen, neon \& magnesium (ONeMg) rather
than carbon and oxygen (CO). This can occur either because the WD
began accreting as an ONeMg WD or because the accretion rate onto the
WD did not allow the WD to remain a CO WD (e.g. Nomoto \& Iben 1985;
Nomoto \& Kondo 1991; Martin et al.\ 2006). Such WDs will produce a
neutron star (NS) via AIC.  Currently it does not appear likely that
AIC produces sufficiently large kicks to disrupt such close binaries
(see, e.g., Podsiadlowski et al.\ 2004).

\subsubsection{Alternatives to SN Ia: White-dwarf mergers?}

Single LMWDs may be the product of the merger of two low-mass He WDs,
with formation rates comparable to or greater than the SN Ia rate
(see, e.g., Han et al.\ [2002] and references therein). However, Han
et al.\ predict masses in excess of $\rm 0.4~M_{\odot}$.

\section{Single-Degenerate SN Ia Populations}
\label{sec:SNIaPops}

\subsection{Expected formation channels}
\label{sec:formation}

We do not present an exhaustive description of the full
evolutionary histories for single-denegerate SN Ia progenitors (see,
e.g. Whelan \& Iben 1973; Nomoto 1982; van den Heuvel et al.\ 1992;
Rappaport, Di Stefano \& Smith 1994; Hachisu, Kato \& Nomoto 1996,
1999; Li \& van den Heuvel 1997; Langer et al.\ 2000; Hachisu \& Kato
2001; Han \& Podsiadlowski 2004).  There is no clear consensus on
which donor stars are likely to produce a type Ia supernova. The
favoured options involve either donors on the main sequence (MS) or
the subgiant branch (known as the supersoft channel), or red-giant
(RG) donors.

While the supersoft channel (e.g. Han \& Podsiadlowski 2004) is
arguably the favoured channel for the majority of SNe Ia, Hachisu,
Kato \& Nomoto (1996, 1999) and Hachisu \& Kato (2001) suggest
situations in which a low-mass giant star may take a WD to the
Chandrasekhar mass $M_{\rm Ch}$ at long orbital periods.\footnote{The
point at which an explosive nuclear runaway occurs in a non-rotating
CO WD is slightly below the Chandrasekhar mass: Nomoto, Thielemann \&
Yokoi (1984) calculated a mass of $\rm \sim 1.378 M_{\odot}$.}
Sokoloski et al.\ (2006) used the 2006 outburst of RS Ophiuchi to
confirm the conclusions of Hachisu \& Kato by inferring that RS Oph
contains a very massive WD ($M_{\rm WD} \approx 1.4 ~{\rm
M_{\odot}}$). It is worth noting that we cannot be sure that RS Oph
contains a CO WD rather than an ONeMg one and so we cannot be sure
that it will explode rather than collapse. Observational support for a
giant donor in a system which produced a SN Ia has been provided via
the observations by Patat et al.\ (2007) of SN 2006X.

King, Rolfe \& Schenker (2003) have also suggested that an accreting
WD may not reach $M_{\rm Ch}$ via the supersoft channel alone, but
that a later phase of WD growth could occur in long-period dwarf
novae. They argue that, even though the average mass-transfer rate
does not reach the steady-burning band (Paczy\'{n}ski \& \.{Z}ytkow
1978; Nomoto \& Kondo 1991), the accretion rate may be high enough for
the WD to grow during dwarf nova outbursts driven by the
thermal-viscous disc instability (Cannizzo, Ghosh \& Wheeler
1982). Providing the correct mass-accretion rate for the CO WD to grow
to $M_{\rm Ch}$ is a significant uncertainty in all these models.

\subsection{Remnant mass}
\label{sec:RemMass}

In order to understand the formation of LMWDs in systems which produce
SN Ia explosions, we must consider the mass and evolutionary stage of
the donor star at the point of the explosion and also the extent to
which the donor loses mass because of the explosion. There is a clear
division between pre-giant and giant donor stars, with giant donors
apparently able to leave LMWD remnants.

Marietta, Burrows \& Fryxell (2000) performed numerical simulations of
the effect of a SN Ia explosion on the companion star. They found that
$\rm 0.15$ to $\rm 0.17 ~{\rm M_{\odot}}$ is stripped away from a $\rm
1 ~{\rm M_{\odot}}$ main-sequence or subgiant companion by the
high-velocity ejecta. Han \& Podsiadlowski (2004) found in their
population synthesis simulations of the supersoft channel that, at the
time of the explosion, the companion has a mass between $\rm \gtrsim
0.5 ~{\rm M_{\odot}}$ and $\rm 2.2 ~{\rm M_{\odot}}$, with a typical
mass of $\rm 1 ~{\rm M_{\odot}}$ (for more details see also Han
2008). Applying the results of Marietta et al.\ as a percentage --
15\% of the donor mass -- leads to a lowest estimated remnant mass of
$\rm \approx 0.42~{\rm M_{\odot}}$. If the WD explodes as it reaches
$M_{\rm{Ch}}$, then this remant mass is a lower limit for the MS
channel, assuming negligible subsequent mass loss in a wind. Hachisu
\& Kato (2001) found a lower limit on the mass of the donor from the
supersoft channel (at the time of the SN) of $\rm > 1.3 ~{\rm
M_{\odot}}$ (assuming an initial white dwarf mass of $\rm 1 ~{\rm
M_{\odot}}$). Despite these differences, both studies suggest that it
is difficult to produce LMWDs via main-sequence or subgiant donors.

Marietta et al.\ also found that a red-giant donor will lose almost
its entire envelope (96\% -- 98\%) due to the impact of the SN Ia
explosion and leave only the core of the star, providing a possible
pathway for the formation of a subset of single, low-mass He WDs. For
the RG channel, Hachisu \& Kato (2001) found a lower limit on the
total donor mass of $\rm \gtrsim 0.4 ~{\rm M_{\odot}}$. If the RG
channel produces SNe Ia, then ram-pressure stripping of the donor's
envelope would be expected to lead to the formation of LMWDs. The
remnant WD mass is dependent on the core mass of the donor at
explosion and is therefore strongly correlated with the orbital period
(see section \ref{sec:OrbitalVelocities}).

One formation channel that is rarely discussed in the literature is
one where the donor star is a hot subdwarf star (see, e.g., Geier et
al.\ 2007). We do not expect significant stripping of the donor by the
supernova ejecta in this case, as the donor star will be tightly
bound, but the mass of the donor star could easily be low enough for a
single LMWD to be formed by the natural evolution of the donor star. 

\subsection{Sub-Chandrasekhar mass explosions}
\label{sec:subCh}

A variation on the above models for SN Ia involves the explosion of
sub-$M_{\rm Ch}$ CO WDs covered with a thick helium layer (Woosley \&
Weaver 1994). In that model, the detonation of the helium layer is
responsible for triggering the supernova. Fink, Hillebrandt \&
R\"{o}pke (2007) found that sub-$M_{\rm Ch}$ explosions were unlikely
to be able to explain either normal or subluminous SN Ia, but there
could be implications for our LMWD formation channel if a significant
fraction of SNe Ia were found to be produced by sub-$M_{\rm Ch}$
detonations.\footnote{If such sub-$M_{\rm Ch}$ explosions occur
but are {\emph{not}} seen as SN Ia, it is not clear to us whether such
an event would remove the RG donor's envelope in order to produce a
LMWD.} Qualitatively there would be little change to our model, as
the RG donors would still be stripped of their envelopes and produce
LMWDs. The quantitative remnant mass distribution may be
different. For example, these sub-$M_{\rm Ch}$ explosions might
plausibly happen when the donor stars are lower on the giant branch
than for the standard model. In that case, the typical remnant WDs may
be less massive, the orbital periods at explosion lower and the
runaway remnant velocities higher than for detonations at $M_{\rm
Ch}$. The remnant velocity is considered hereafter as a diagnostic of
the orbital period at explosion, assuming Chandrasekhar-mass
explosions.

\subsection{Binary orbits and runaway velocities}
\label{sec:OrbitalVelocities}

If single LMWDs have been released from binary systems in which the
other component has exploded as a type Ia supernova, the space
velocity of the remnant should be a useful diagnostic of the orbital
period at explosion. As our arguments above suggest that the LMWDs are
most likely to originate in systems with red-giant donor stars, the
relationship between core mass and orbital period in such systems can
act as a further constraint.

In what follows we assume that the donor stars are filling their
Roche lobes, as it seems to us that the mass transfer in systems which
produce SN Ia is most likely to be due to Roche-lobe
overflow. However, it is not known whether this is the case as, for
example, it is unclear whether the donor star in RS Ophiuchi is
rotating synchronously with the binary orbit (see, e.g. Murset \&
Schmid 1999; Zamanov et al. 2007). If the donor stars do not fill
their Roche lobes, then the following method may slightly overestimate
the runaway velocities for a given remnant mass.

If we define $q$ as $M_{\rm 2}/M_{\rm 1}$ (where $M_{1}$ is the mass
of the SN Ia progenitor and $M_2$ is the mass of the companion
producing the LMWD), write the total mass of the system (in solar
units) as $M_{\rm tot,\odot}$ and the pre-SN orbital period (in days)
as $P_{\rm days}$, we can write the pre-SN orbital
velocity as:
\begin{equation}
\label{eq:Vorb}
V_{\rm orb} \approx \frac{213}{1+q} \left( \frac{M_{\rm tot,\odot}}{P_{\rm days}} \right)^{1/3} ~{\rm km/s}.
\end{equation}
This shows that, for $M_{\rm 2} \ll M_{\rm 1}$, the orbital
velocity is relatively insensitive to the donor mass.

Furthermore, assuming that the remnant WD mass equates to the core mass of
the donor at the time of the SN explosion, then the well-defined
relationship between core mass and radius for giant-branch donors
leads to an expression for the orbital period of the system at the
supernova stage. Rappaport et al.\ (1995) found the period-mass relation:
\begin{equation} 
\label{eq:McP}
P_{\rm orb} \approx 0.374 \left( \frac{R_{\rm 0} M_{\rm wd}^{4.5}}{1 + 4 M_{\rm wd}^4} +
0.5 \right)^{3/2} M_{\rm wd}^{-1/2} ~{\rm days},
\end{equation}
where $M_{\rm wd}$ is the mass of the future WD (currently the core of
the giant star) in units of solar masses. Their preferred value for
the fitting parameter $R_{\rm 0}$ was $\rm 4950~ R_{\odot}$, which we
also adopt. If we now assume a primary mass of 1.4 $\rm ~M_{\odot}$
and a total secondary mass at the time of explosion of 0.5 $\rm
~M_{\odot}$, we obtain for a range of core masses:
\begin{tabbing}

$P_{\rm orb}({\rm 0.2 ~M_{\odot}}) = {\rm 6.74~d}~~~~$ \= $\rm ~~~\Rightarrow~~~$ \= $V_{\rm orb}={\rm 103~km~s^{-1}}$ \\

$P_{\rm orb}({\rm 0.25 ~M_{\odot}}) = {\rm 23.7~d}$ \> $\rm ~~~\Rightarrow~~~$ \> $V_{\rm orb}={\rm 68~km~s^{-1}}$ \\

$P_{\rm orb}({\rm 0.3 ~M_{\odot}}) = {\rm 69.4~d}$ \> $\rm ~~~\Rightarrow~~~$ \> $V_{\rm orb}={\rm 47~km~s^{-1}}$ \\

$P_{\rm orb}({\rm 0.35 ~M_{\odot}}) = {\rm 171.8~d}$ \> $\rm ~~~\Rightarrow~~~$ \> $V_{\rm orb}={\rm 35~km~s^{-1}}$ 
\end{tabbing}
where $V_{\rm orb}$ is the total orbital velocity. Using the same
component masses, we record for comparison that $P_{\rm orb} = {\rm
1~d}$ corresponds to $V_{\rm orb} \approx {\rm 194~km~s^{-1}}$ and
$P_{\rm orb} = {\rm 1~h}$ to $V_{\rm orb} \approx {\rm 560~
km~s^{-1}}$. The orbital velocities at explosion in the simulations of
Han \& Podsiadlowski (2004) range from 80--230 km s$^{-1}$ for MS and
subgiant donors.

RS Ophiuchi has an orbital period of $\approx$457 d. Inverting equation
\ref{eq:McP} above, this corresponds to a core mass for the donor of
slightly over $\rm 0.4 ~M_{\odot}$, still within the mass range for a
LMWD should the envelope be removed. 

The arguments in section \ref{sec:RemMass} suggest that LMWDs are
produced by giant donors, as long as the orbital period is not so long
that the core has already grown to 0.5 $\rm {\rm M_{\odot}}$ by the
time their tenuous envelopes are stripped by the supernovae
ejecta. So, in contrast to the high-velocity WDs observed by
Oppenheimer et al.\ (2001) and interpreted by Hansen (2003) as
remnants of SNe Ia with main-sequence donors, it would be consistent
to find that the single LMWD population was not significantly
kinematically heated.

In section \ref{sec:RemMass} we identified two potential SN Ia
formation channels able to produce LMWDs. We note that whilst one of
those sets of donor stars (red giants) would leave LMWD remnants with
low runaway velocities, the other (hot subdwarfs) would result in
high-velocity LMWDs.

\section{Single UCWDs as LMWDs \& SN Ia remnants?}
\label{sec:UCWDs}

We have argued that single LMWDs can be produced from
single-degenerate SNe Ia with red-giant donors. Single LMWDs are
inferred to exist and the most natural explanation, especially for the
lower-mass LMWDs, seems to be that some single-degenerate SN Ia occur
with red giant donors. However, there is no obvious collected sample
of LMWDs to examine as potential SN Ia remnants. Independently of our
arguments above, it may well be that the known set of apparently
single UCWDs constitutes or contains a useful sample of single LMWDs.
When a suitable sample of single LMWDs becomes available our work
should be extended.

\begin{figure}
\begin{centering}
\epsfig{file=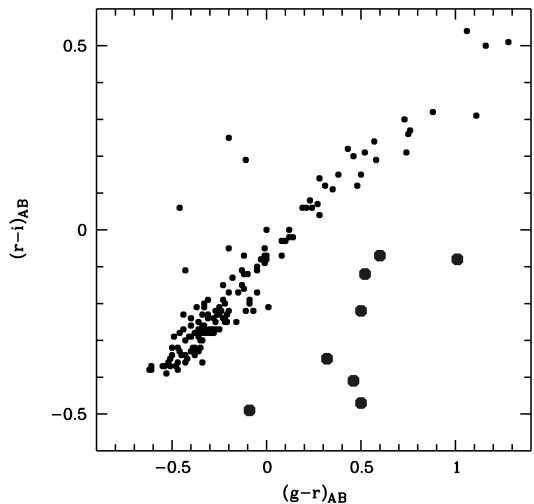, width=7cm}
\caption{
\label{fig:UCWDs}
A colour-colour diagram of white dwarfs. The known ultra-cool white
dwarfs are represented by large circles, while the Kleinman et
al.\ (2004) sample of normal WDs are shown as small circles. Table
\ref{tab:UCWD_dist} lists the UCWDs plotted here along with the
relevant references. (Adapted from Wolf [2005].)}
\end{centering}
\end{figure}

\begin{table*}
\centering
\caption{Distances and tangential velocities of the UCWD sample
(as in Fig. \ref{fig:UCWDs}). For all objects except LHS
3250, the distance is inferred from an assumed absolute magnitude. Our
preferred measure assumes LHS 3250 is a typical UCWD; we also present
alternative values based on the range of absolute magnitudes assumed in the
literature.
\label{tab:UCWD_dist}}
\begin{tabular}{lccccc}
\hline
\hline
Name & \multicolumn{2}{c}{ Estimated properties for $M_{\rm V}$=15.7$~^{a}$} &$~$& \multicolumn{2}{c}{ Estimated properties for $M_{\rm V}$=16.5 $\pm$1.0$~^{b}$ } \\
     & Distance (pc) & Tangential velocity (km/s) &$~$& Distance (pc) & Tangential velocity (km/s) \\
\hline
LHS 3250$~^{1}$       & 30 & 81  &$~$& \multicolumn{2}{c}{LHS 3250 has a known parallax ~~~~~~~~~}  \\
LHS 1402$~^{2}$       &$~$ 24$~^{c}$ & 56  &$~$&$~$ 12--31$~^{d}$ & 28--72 \\
SDSS J0947$~^{3}$     & 47           & 18  &$~$& 21--52           & 8--21  \\
SDSS J1001$~^{3}$     & 64           & 107 &$~$& 28--71           & 47--119  \\
SDSS J1220$~^{3}$     & 64           & 154 &$~$& 28--71           & 68--170  \\ 
SDSS J1337$~^{4}$     & 54           & 46  &$~$& 23--58           & 20--49  \\
SDSS J1403$~^{3}$     & 44           & 60  &$~$& 19--49           & 26--66  \\
COMBO-17 J1143$~^{5}$ & 169          & 42  &$~$& 37--188          & 9--46 \\
\hline
\hline
\multicolumn{6}{l}{$^{a}$: {\footnotesize i.e. taking LHS 3250, with a known trigonometric parallax, as representative of all UCWDs.}} \\
\multicolumn{6}{l}{$^{b}$: {\footnotesize i.e. using a conservative assumption adopted by Salim et al.\ (2004), Gates et al.\ (2004) and Wolf (2005).}} \\
\multicolumn{6}{l}{$^{c}$: {\footnotesize For LHS 1402, we compare a B magnitude of 18.32 (Oppenheimer et al.\ 2001) with LHS 3250 (18.85; Harris et al.\ 1999).}} \\
\multicolumn{6}{l}{$^{d}$: {\footnotesize Correcting for different photometry, we compare a B magnitude of 18.2 with the g magnitude (20.04) of SDSS 1001.}} \\
\multicolumn{6}{l}{$^{1}$: {{\footnotesize Harris et al.\ 1999}} $~~^{2}$: {{\footnotesize Oppenheimer et al.\ 2001 }}  $~~^{3}$: {{\footnotesize Gates et al.\ 2004 }} $~~^{4}$: {{\footnotesize Harris et al.\ 2001}} $~~^{5}$: {{\footnotesize Wolf 2005}} } \\
\hline
\hline
\end{tabular}
\end{table*}

\subsection{The UCWD sample}
\label{sec:UCWDsample}

In selecting sub-samples of the WD population, UCWDs (see, e.g. Harris
et al.\ 1999 \& 2001; Gates et al.\ 2004; Wolf 2005) are clear
outliers in a colour-colour diagram (see Fig. \ref{fig:UCWDs}). Their
optical colours distinctly separate them from the normal WD
population, and they are an interesting curiosity in appearing to
become bluer as they cool down, possibly due to the effects of
collisionally induced absorption (CIA) by hydrogen molecules in the
atmosphere (Bergeron et al.\ 1994; see also Kowalski \& Saumon
2006). When CIA affects only the infrared part of the spectrum, WDs
are classed as {\emph{cool}}; if CIA also affects the optical colours,
then the WD is admitted into the select group of UCWDs (see Fig. 1;
also Wolf 2005). The transition temperature between {\emph{cool}} and
{\emph{ultra-cool}} is $\rm \lessapprox 4000 K$.

The known UCWDs constitute a clean observational sample; they are easy
to identify, and with such low luminosities (notably in the
ultra-violet and near infra-red) and line-free spectra, it would be
hard to hide a light-emitting close companion that is anything but
another UCWD (see section \ref{sec:UCWDsingle}).  Table
\ref{tab:UCWD_dist} contains the UCWD sample we use. The estimates of
the tangential velocities for these objects depend on their assumed
distances, and Table \ref{tab:UCWD_dist} shows the range of velocities
obtained for assumptions taken from the literature. For that sample we
shall adopt the tangential velocities obtained by taking the absolute
magnitude of LHS 3250 (which is the only UCWD with a known parallax)
to be representative of the whole sample.

Since we first began this work, a new sample of twenty-four UCWD
\emph{candidates} has been presented by Vidrih et al.\ (2007). We are
not convinced that they are cold enough to conform to our strict
criteria as UCWDs, which may mean that this sample is more likely to
be contaminated by non-LMWD objects (see section 4.2). However, in our
later figures we shall show for comparison this new, independent,
sample alongside the smaller set from Table
\ref{tab:UCWD_dist}.\footnote{We have not included an extremely recent
sample of seven very cool white dwarfs (Harris et al.\ 2008).}

\subsection{Are the observed UCWDs mostly LMWDs?}
\label{sec:UCWDsLMWDs}

The one UCWD with a known parallax (LHS 3250) has an absolute
magnitude of $M_{\rm V}=15.72$. This is brighter than expected for
anything other than a LMWD; hence Bergeron \& Leggett (2002) conclude
that the mass of LHS 3250 is $\rm 0.23~M_{\odot}$.\footnote{Harris et
al.\ (2001) also suggest that LHS 3250 is probably a low-mass helium
core WD based on its absolute magnitude; however, the earlier paper by
Harris et al.\ (1999) noted that a pair of UCWDs in a binary might
help explain the higher luminosity without recourse to a low-mass WD.}
Unfortunately we do not have such good mass estimates for all single
UCWDs.

The argument that the observed UCWDs are mostly LMWDs is partly built
upon theoretical WD cooling tracks. Whereas a $\rm 0.6 ~M_{\odot}$ WD
takes more than $\rm 9~Gyr$ to cool to 4000 K ({\emph{after}} the
formation of the WD), the $\rm 0.3~M_{\odot}$ WD of equivalent
composition takes less than $\rm 4~Gyr$ (Bergeron, Leggett \& Ruiz
2001; Bergeron et al.\ 2005). As the cooling of WDs is a function of
composition it is likely that not all UCWDs are LMWDs (see, e.g.,
Hansen 1999, who requires WD masses $\rm\lesssim 0.25~M_{\odot}$ in
order for those objects to cool to 4000K within 7 Gyr).\footnote{We
note in advance that this mass of $\rm\lesssim 0.25~M_{\odot}$ is in
good agreement with our results in section \ref{sec:PopKinRes}
combined with the core-mass orbital-period relation; it is thus
self-consistent.} Extremely low-mass WDs ($\lesssim 0.17 {\rm
M_{\odot}}$) seem to cool more \emph{slowly} than more massive WDs,
due to the retention of a relatively thick hydrogen envelope
(e.g. Panei et al.\ 2007). It is not clear whether the LMWD remnants
that have been formed by having their envelopes forcibly removed by a
supernova shockwave will retain a thick hydrogen envelope.

Hence we expect that the observed, apparently single, UCWDs are
dominated by single LMWDs if they exist, partly as they are
significantly more luminous than massive WDs and hence more likely to
be discovered.  Although it is unfortunate that we cannot prove what
fraction of single UCWDs are LMWDs, in section \ref{sec:PopNumb} we
show that the observed numbers of single UCWDs could {\emph{all}} be
single LMWDs produced via a SN Ia explosion.

\subsection{Are the apparently single UCWDs mostly single?}
\label{sec:UCWDsingle}

Seven of the eight UCWDs in Fig. 1 have no known companion. The
exception -- SDSS J0947 -- has a common proper motion companion (Gates
et al.\ 2004). The 20 arcsecond angular separation of J0947 from its
potential companion implies a projected separation of over ${\rm 2
\times 10^{5}~ R_{\odot}}$ for a distance of 47 pc; if that companion
really forms a binary with J0947 (rather than being chance projection), it
could not have influenced the evolution of the progenitor of the J0947
UCWD.

We do, however, need to consider whether these apparently single UCWDs
really are single. We fully expect that UCWDs should exist in binary
systems, but in the following we argue that these are unlikely to
contaminate our sample; in many cases, a companion would even
completely hide a UCWD.

\subsubsection{Non-degenerate companions}

An M-dwarf with an absolute V magnitude of $\sim 16$ -- similar to the
UCWD LHS 3250 -- would have a mass of $\rm \sim 0.1~{\rm M_{\odot}}$
(Delfosse et al.\ 2000). Such a star would be bright in the infra-red,
where the emission of UCWDs is strongly suppressed.\footnote{For
example, a $\rm \sim 0.1~{\rm M_{\odot}}$ star is about six magnitudes
brighter in the J band than the V band (Delfosse et al.\ 2000), wheras
LHS 3250 and SDSS 1337 are dimmer in the J than V bands (Harris et
al. 2001). This trend is greater at longer wavelengths.} Furthermore,
M-dwarfs are rich in spectral lines, so could be easily detected.

\subsubsection{White-dwarf companions}

A binary containing a non-ultra-cool WD should be identifiable: an
advantage of UCWDs is that their low luminosity and featureless
spectrum makes it hard for them to possess an undetectable hotter WD
companion.  Their characteristic spectral energy distributions mean
that a non-UCWD companion would be brighter than the UCWD in either
near infra-red or ultra-violet light.  However, a spectroscopic
UCWD-UCWD binary would be difficult to distinguish from a single
UCWD. We have no reason for thinking that such binary UCWDs do not
exist, and Harris et al.\ (2008) have discovered one system which
might eventually be expected to become such a binary. Although we do
not expect that they are common enough to dominate the population, the
possibility that the apparently single UCWDs have extremely cool or
faint WD companions should be studied further.

\subsubsection{Neutron-star companions}

From the arguments in section 2.1 it is clear that, in order to make a
LMWD in a system where the NS was produced in a core-collapse
supernova, the LMWD progenitor must lose mass after the formation of
the NS. Hence, either mass transfer onto the NS or a common-envelope
phase (Paczy\'nski 1976) would be required. In the former case, the
pulsar is expected to be recycled into a millisecond pulsar
(MSP).

It is thus reasonable to expect that UCWDs might be seen as companions
to millisecond pulsars. This has been claimed for PSR J0751+1807
(Bassa, van Kerkwijk \& Kulkarni 2006a), although the LMWD in that
system is not quite formally ultra-cool. The nearby millisecond pulsar
J0437-4715 also seems to be accompanied by a borderline UCWD
(Danziger, Baade \& Della Valle 1993). 
It could be argued that, since we do see LMWDs with pulsars, it is not
a surprise that some of these LMWDs happen to be in systems where we
do not see the pulsar (see also van Kerkwijk, Bergeron \& Kulkarni
1996; van Kerkwijk et al.\ 2000; Bassa et al.\ 2006b). However, the
space densities of these different samples are extremely
different. The companions to pulsars have been discovered
{\emph{because}} they are orbiting a pulsar, unlike the
photometrically-discovered UCWD sample: Bassa, van Kerkwijk \&
Kulkarni state that PSR J0751+1807 is $\sim$0.6 kpc away, considerably
further than any of our UCWD sample (see Table \ref{tab:UCWD_dist}),
and PSR J0437-4715 is 150 pc away (Danziger et al.\ 1993), beyond all
but one of the photometrically-discovered UCWDs.

A crude comparison between the possible detection volumes of the
single UCWDs and the pulsar companions gives $\sim(600/70)^{3} \approx
630$. Here we have considered that COMBO-17 J1143, as a
serendipitously discovered object, does not give a good measure of the
systematic detection volume of the UCWDs in our sample. This suggests
that, even if the companions to PSR J0751+1807 and PSR J0437-4715 were
cool enough to be UCWDs, the seven systematically discovered UCWDs in
our sample are $\sim2200$ times more abundant than those with pulsar
companions. This factor can be increased by another order of magnitude
if one assumes that all the known nearby pulsars have been studied,
but that only a tenth of the sky has been surveyed to the same depth
as the field that produced the SDSS UCWD sample.
To only observe one pulsar in $\sim 20,000$ supposedly MSP-containing
objects would imply extremely narrow-beam pulsar emission. Unless
those unseen MSPs were somehow unusual, this would suggest that many
more pulsars exist than we currently expect, significantly worsening
any mismatch between the inferred birthrates of LMXBs and MSPs
(e.g. Kulkarni \& Narayan 1988; Lorimer 1995; Pfahl, Rappaport \&
Podsiadlowski 2003).\footnote{If we repeat this estimate for the new
Vidrih et al.\ sample of 24 UCWDs, which extends to a distance of 180
pc but over only 250 square degrees of the sky, then we have $\sim
(600/180)^{3} \times (24/2) \times (41523/250)$, i.e. single UCWDs
outnumber those with MSPs by a factor of over 70,000 (over 27,000 if
UCWDs are twice as bright as Vidrih et al.\ assume and their survey
depth extends to 250 pc).}

If a NS was formed in the system through AIC and was not subsequently
spun up, it could reasonably be expected not to emit pulsar
radiation. A black-hole companion would, of course, not be expected to
emit pulsar radiation, but a local space density for such black-hole
binaries of $\rm \sim 10^{-5} ~pc^{-3}$ (see section
\ref{sec:PopNumb}) is highly unexpected (see, e.g., Romani 1998).

\begin{figure*}
\begin{centering}
\mbox{
\epsfig{file=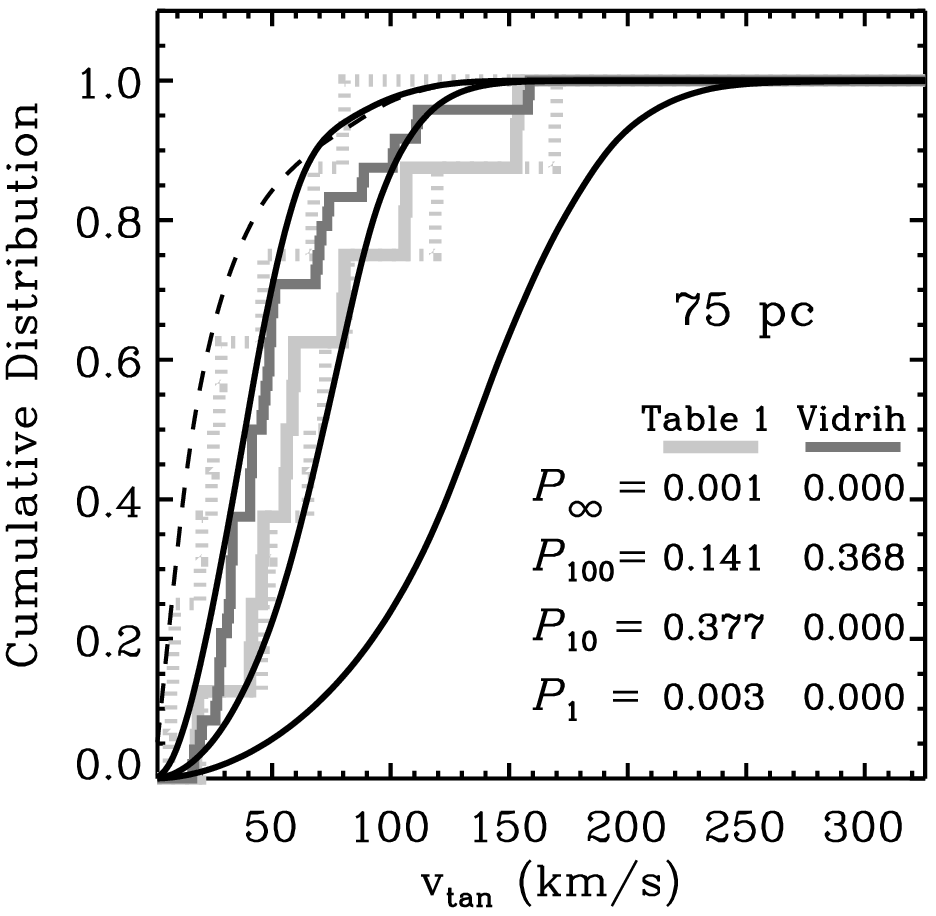, height=6.5cm}
\epsfig{file=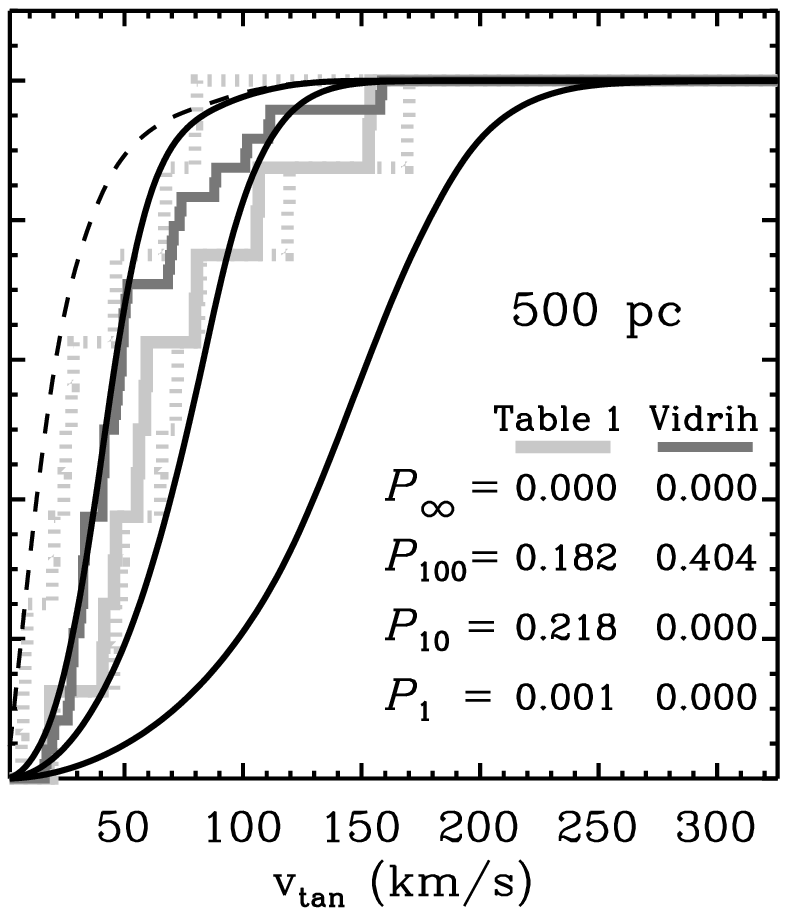, height=6.5cm}
\epsfig{file=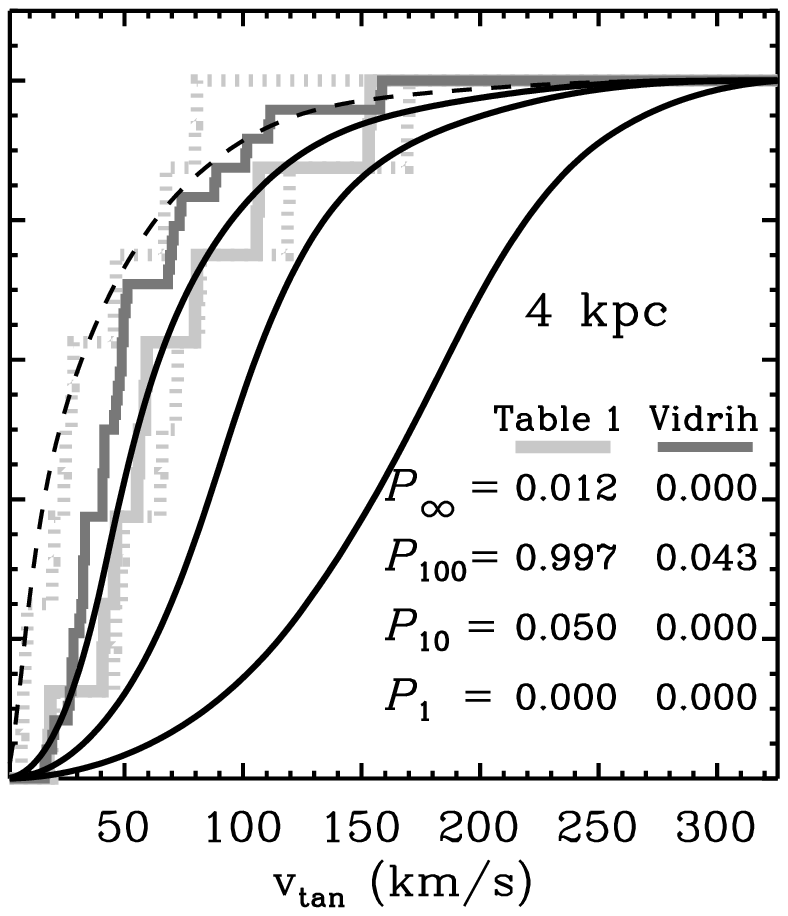, height=6.5cm}
}
\caption{
\label{fig:CLOUDS}
Comparison of the tangential velocities resulting from our Galactic
integrations (smooth black curves), with the tangential velocities of
the observed UCWDs (grey step functions). The solid light grey step
function uses the distance estimates from the assumption that LHS 3250
is representative of the UCWD population, and the broken grey step
functions encompass the wider range of distances in Table
\ref{tab:UCWD_dist}. The dark grey step function uses the new sample
of UCWD candidates from Vidrih et al.\ (2007), adopting their
assumptions for the UCWD distances. The solid black curves in each
panel are for orbital periods at the time of the explosion of 100, 10
and 1 days (left to right). The dashed black curve represents a
population which receives no `kick' due to the break-up of a binary
system. The Kolmogorov-Smirnov acceptance probabilities for the
individual models are given in each panel, compared to both the
objects in Table \ref{tab:UCWD_dist} and the Vidrih et al.\
sample. Here $P_{\rm \infty}$ refers to the model with no added `kick'
velocity from binary break-up and $P_{\rm 100}$, $P_{\rm 10}$ and
$P_{\rm 1}$ to the curves representing the 100, 10 and 1 day
orbital-period populations, respectively (see text). The simulations
assumed initial disc scale-heights of 75 pc, 500 pc and 4 kpc (left to
right). If these objects were once the cores of red-giant donor stars
in SN Ia producing systems, the simplest expectation would be for
orbital periods of $\sim$100 d (see text). }
\end{centering}
\end{figure*}

\begin{figure*}
\begin{centering}
\mbox{
\epsfig{file=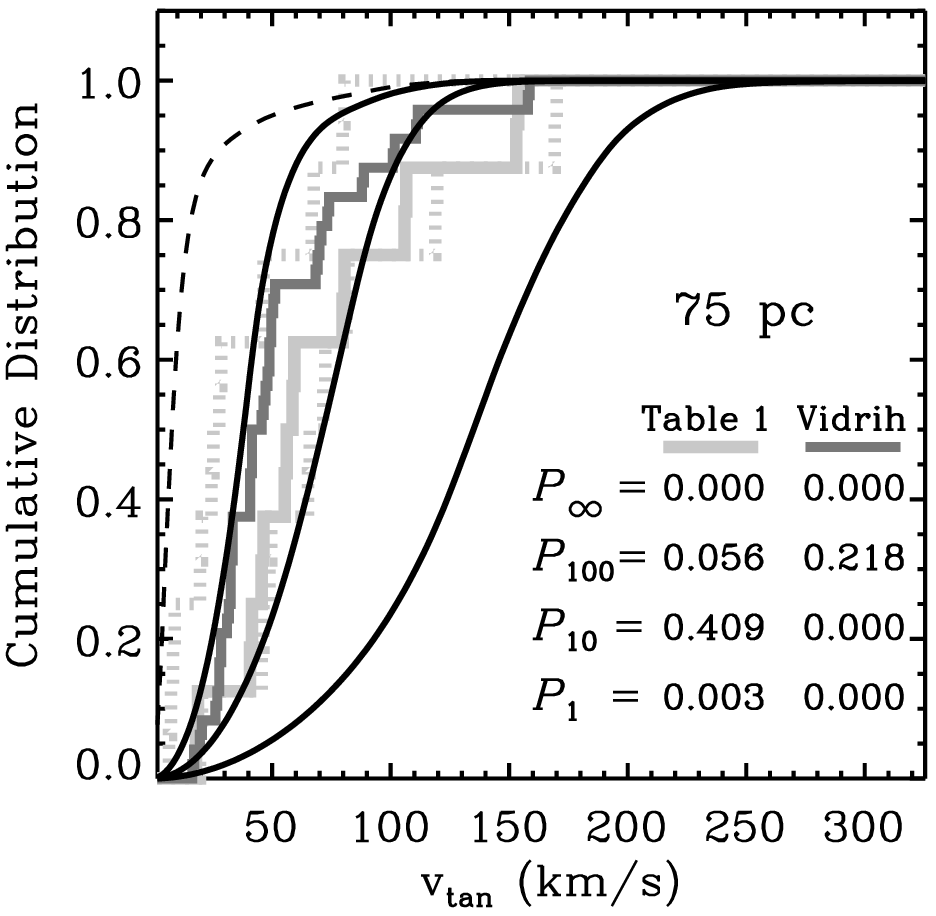, height=6.5cm}
\epsfig{file=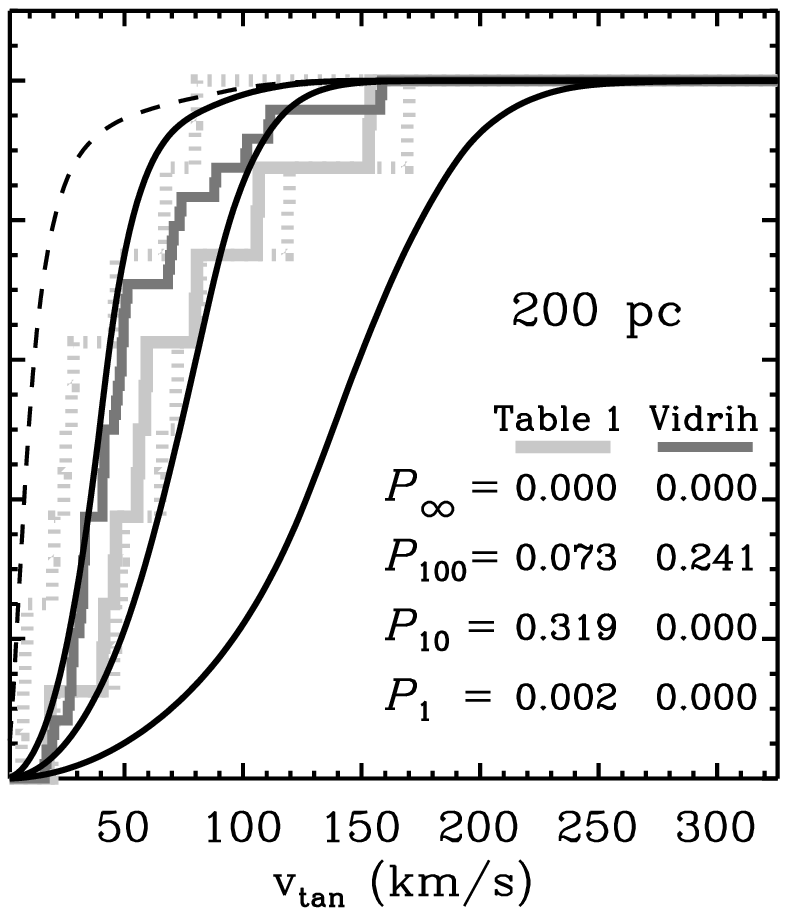, height=6.5cm}
\epsfig{file=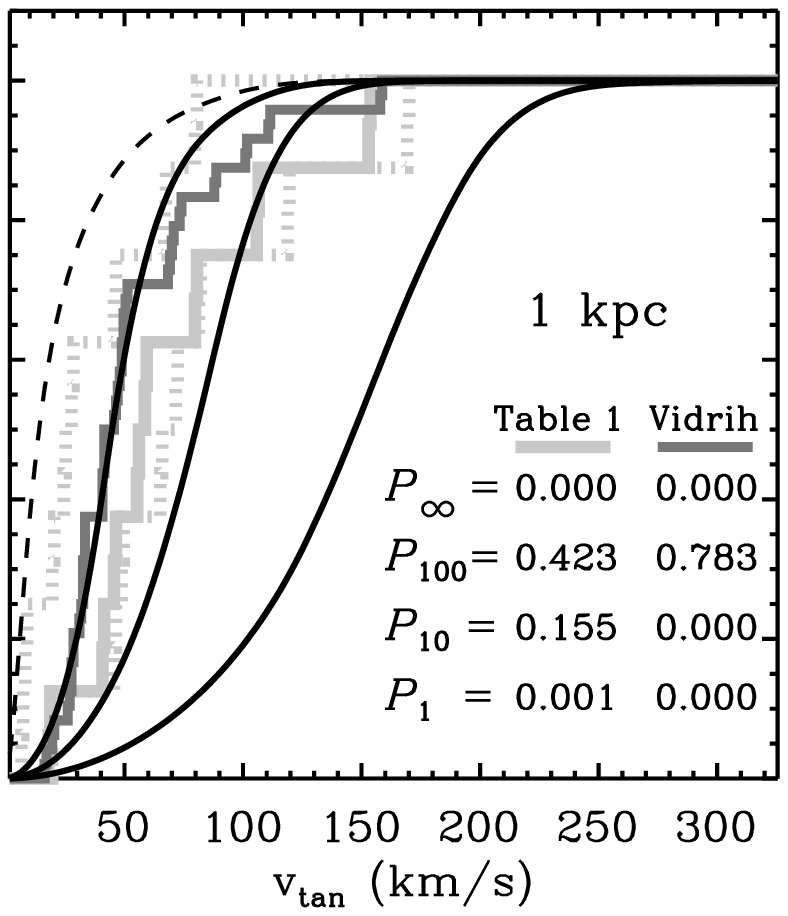, height=6.5cm}
}
\caption{
\label{fig:NOCLOUDS}
As Fig. \ref{fig:CLOUDS}, for simulations that do not include any
scattering from giant molecular clouds. The plots assume initial disc
scale-heights of 75 pc, 200 pc and 1000 pc (left to right).  }
\end{centering}
\end{figure*}

\subsection[]{UCWD population numbers}
\label{sec:PopNumb}

Gates et al.\ (2004) estimated a space density for UCWDs of $\rm 3
\times 10^{-5} ~pc^{-3}$ from a sample of 6 objects found in the Sloan
digital sky survey.\footnote{For the Viridh et al.\ sample, we
estimate a space density of $\sim 16 \times 10^{-5} ~pc^{-3}$ for a
survey depth of 180 pc, and $\sim 6 \times 10^{-5} ~pc^{-3}$ for a
survey depth of 250 pc.} This rough figure does compare with an
estimate of the SN Ia rate integrated over time and space. The local
stellar density of $\rm 0.1 ~{\rm M_{\odot}} ~pc^{-3}$ (Binney \&
Merrifield 1998), combined with a mass for the thin disc of $\rm \sim
4 \times 10^{10} ~{\rm M_{\odot}}$ and the assumption that the mass
fraction of UCWDs is constant throughout the disc, produces an
estimate for the number of UCWDs of $\sim 10^{7}$.

We approximate the current SN Ia rate in the Galactic disc as $\rm
\sim 7 \times 10^{-3} ~/ yr$ (using the same disc mass as above and
the SN Ia rate per unit mass of Mannucci et al.\ 2005). Multiplying
this rate by a Galactic age of $\rm \sim 10~Gyr$ leads to an estimate
of a total of $\rm \sim 7 \times 10^{7}$ remnants.\footnote{The
supernova rate may well have been different in the past. Hansen (2003)
made estimates for the number of SN Ia remnants in the Milky Way based
on the amount of iron in the Galaxy and found a range between $\rm 4
\times 10^{7}$ and $\rm 2.2 \times 10^{8}$ objects.}

Even if we halve this number of remnants to allow for the WD cooling
time (see section \ref{sec:UCWDsLMWDs}), the number of single UCWDs is
easily consistent with them being produced through SN Ia. Indeed,
these numbers suggest that only a subset of SNe Ia produces single
UCWDs, which is as expected if only a subset of the SN Ia formation
channels can produce LMWDs (see sections \ref{sec:formation} \&
\ref{sec:RemMass}).

\subsection[]{Population kinematics: Method}
\label{sec:PopKin}

The lack of lines in UCWD spectra means that we do not know the radial
velocities for our sample. However, we can examine the population
kinematics using only the information from the tangential velocities. We now
investigate whether their observed space velocities are consistent
with single LMWDs released from binaries with a range of orbital
periods.

For a range of inital parameters, we integrated the motion of $\rm
10^{5}$ assumed SN Ia remnants for up to 10 Gyr through the Galactic
potential (using a similar procedure to Brandt \& Podsiadlowski 1995),
orientating the orbital velocity vector of the donor at random when
the binary is disrupted. For each integrated population, we used a
single value of orbital period (and hence orbital velocity) at the
time of the explosion.\footnote{We expect that in reality SN Ia
explosions will occur over a range of orbital periods, but the current
data only allow us to investigate broad trends.} For calculating the
orbital velocity at a given orbital period, donor stars are assumed to
be $\rm 0.5 ~{\rm M_{\odot}}$ at the time of the explosion, and the
WDs are assumed to explode at a mass of $\rm 1.4 ~{\rm
M_{\odot}}$. Equation \ref{eq:Vorb} shows that our results should be
relatively insensitive to those assumptions, but in the future we
intend to perform this procedure using the output of our binary
population synthesis calculations.

Each remnant is initially located at random within an axisymmetric
Galaxy modelled by two exponential scale-heights (vertical and
radial). The axisymmetry is also exploited for computational
efficiency: at each integration time-step the view from Earth is
calculated at all points on the solar circle. A further assumption is
that the remnants can be observed to a distance of 160 pc -- broadly
appropriate for UCWDs. Within such a small volume, the space
velocities should only be a very weak function of distance.

The Galactic potential was taken from Paczy\'{n}ski (1990), using the
parameters in Brandt \& Podsiadlowski (1995), with the addition of
scattering from giant molecular clouds (GMCs) randomly distibuted
within the Galactic disc. The total mass in these GMCs is assumed to
be $\rm 1.2 \times 10^{9} ~{\rm M_{\odot}}$, with a mass spectrum
exponent such that ${\rm d}N/{\rm d}\log m \propto m^{-1.7}$ (see, e.g.,
Digel et al.\ 1996; Binney \& Merrifield 1998). 

We do not include any kick imparted by the supernova ejecta in our
simulations, since the simulation of Marietta et al.\ (2000) show that
the kicks due to the supernova interaction (86 km s$^{-1}$ and 49 km
s$^{-1}$ for their main-sequence and subgiant donors, respectively,
with no kick given to the core of the giant donor) are generally small
compared the orbital-velocity kick from the break-up of the binary.

Since the initial disc scale-height of the progenitor population is
uncertain, we present our results for a range of values. The vertical
scale-height of massive stars is 75~pc (van der Kruit 1987), and $\sim
200$~pc is an approximation to a more generic thin disc population
(e.g., Ojha et al.\ 1996, who find a scaleheight of $260\pm 50$~pc for
the Galactic thin disc and $760 \pm 50$ for the thick disc; see also
Kroupa, Tout \& Gilmore 1993). Given previous speculations that UCWDs
are so cool because they are very old objects, we also modelled
initial scale-heights of 500~pc, 1~kpc and 4~kpc.\footnote{Note that
the collective tangential velocities of the sample in Table 1 are
clearly inappropriate for objects from a halo population.}

\subsection{Population kinematics: Results}
\label{sec:PopKinRes}

Figures \ref{fig:CLOUDS} \& \ref{fig:NOCLOUDS} present the results of
our integrations for orbital periods at the time of the supernova of
1, 10 and 100 days, as well as a population which received no
kick. The population which received no kick seems to be difficult to
reconcile with the observed tangential velocities, except for the most
extreme range of luminosities consistent with the literature combined
with an inital vertical scale height of 4 kpc (or greater). The
population released from a one day orbital period also appears
inconsistent with the data, whichever initial scale-height is assumed.
These conclusions are supported by applying the Kolmogorov-Smirnov
test to compare our simulations with the tangential velocity
distribution produced by assuming LHS 3250 is a typical UCWD (Table
\ref{tab:UCWD_dist}). Each plot contains a Kolmogorov-Smirnov
acceptance probability ($P_{x}$) for each curve.\footnote{Here
$1-P_{x}$ gives the probability that the data and model $x$ are drawn
from different distributions, so we can reject a model with $P=0.01$
with 99\% confidence.}

Given the considerable uncertainties, we consider that the extremely
favourable Kolmogorov-Smirnov test for the population released from
100 day orbits with a 4 kpc scale height should not be
over-interpreted, especially as a real SN Ia population would be
expected to have a range of orbital periods.  Both the uncertainty in
the distances to UCWDs and in the formation kinematics restricts our
ability to draw quantitative conclusions. However, both the
observational samples presented in figures \ref{fig:CLOUDS} \&
\ref{fig:NOCLOUDS}, with their different assumptions, seem to suggest
that these apparently single LMWDs have experienced some kick.

The broadly favoured period range at explosion is between 10 and 100
days, with the longer orbital periods preferred for a larger initial
scale-height. The sample presented by Vidrih et al.\ leads to a lower
set of tangential velocities, and their sample strongly prefers 100
over 10 d. This may be because they assume fainter absolute magnitudes
than we adopt, and hence systematically produce lower velocities than
us. Alternatively, their larger but less cool sample may be more likely
to be contaminated by objects which are not single LMWDs. 

This period range of 10--100 d approximately encompasses the core
masses appropriate for the production of LMWDs, exactly as might be
expected if the stellar evolution was truncated by the loss of a giant
star's envelope. This is consistent with an explanation of these UCWDs
as being descended from giant donors. We note, however, that we cannot
exclude a thick disc origin for these objects with an arbitrarily
small kick. Although we cannot use this apparent kick as a definitive
signature of an origin in type Ia supernovae, a thick disc origin
would not falsify our SN Ia hypothesis, but it would suggest longer
orbital periods at the point of explosion.

\section{Discussion}

The main conclusion of this study is that single LMWDs constitute
indirect evidence that SNe Ia are formed through the single-degenerate
channel, specifically from systems with red-giant donors. A field
population of truly single LMWDs, especially with masses less than
$\rm \sim0.4 ~{\rm M_{\odot}}$, would seem to require a SN Ia origin,
or a significant revision of our understanding of either CV evolution
or AIC.\footnote{It may be that explosions of
sub-Chandrasekhar-mass WDs occur which are \emph{not} seen as SN Ia
but \emph{are} capable of stripping the envelopes of RG donor stars.
That would provide a further possible channel for the formation of
single LMWDs.}  That is based on evolutionary arguments (sections 2
\& 3) and does not rely on any kinematic information.

We have simultaneously demonstrated that there is a natural formation
channel for single LMWDs which does not require any unexpected
modification of single-star evolution.

Those conclusions are independent of the nature of apparently single
UCWDs. However the currently observed UCWD sample may be dominated by
LMWDs, and we find that the kinematics of the known single UCWDs seem
to be consistent with an origin in SN Ia with red-giant donor
stars. For reasonably broad assumptions, they even suggest that the
orbital periods at the time of explosion were ${\rm \lesssim
100~d}$. The population of UCWDs require further study in order to
strengthen those specific conclusions.

\subsection{SN Ia progenitor evolution}
\label{sec:ProgEv}

The implication of the existence of single LMWDs is that at least some
single-degenerate SNe Ia occur with the donor star on the giant branch
at the time of the explosion.

Both the high-velocity WD and single LMWD populations seem to contain
members originating in SN Ia explosions. Dwarf star donors at the time
of explosion should acquire higher space velocities than giant donors
(due to their shorter orbital periods), but the less extended donors
should be less stripped by the ram pressure of the explosion (Marietta
et al., 2000) and hence go on to produce higher-mass WDs than a giant
which loses its envelope prior to helium ignition. Hence the
Oppenheimer et al.\ (2001) sample of high-velocity WDs considered by
Hansen (2003) do not need to be low-mass for a SN Ia origin to be
reasonable, nor do single LMWDs need to have high space velocities to
invite a SN Ia explanation for their production. LP 400-22 (Kawka et
al.\ 2006) is notable for being an extreme object using either
selection criterion.

A simplistic comparison of the remnant space densities quoted for the
Oppenheimer et al.\ (2001) high-velocity sample ($\rm 1.8 \times
10^{-4}~pc^{-3}$) and the Gates et al.\ (2004) UCWDs ($\rm \sim3
\times 10^{-5}~pc^{-3}$) suggests that the supersoft channel is almost
an order of magnitude more important than the red-giant channel,
though presumably some single LMWDs destined to be single UCWDs have
not yet cooled sufficiently to become UCWDs, and some UCWDs may not be
LMWDs. The relative importance of these formation channels will be an
interesting quantity to constrain with future data and compare with
binary population synthesis models. Perhaps more decisively, the
estimates of Hansen (2003) and section \ref{sec:PopNumb} imply that
the total number of observed remnants is consistent with the expected
number of past SNe Ia in our Galaxy. If this is confirmed by future
work, it would leave room for only a minority of SNe Ia to result from
double-degenerate systems.

One curiosity of our study of UCWDs is that the range of orbital
periods we infer for the SN Ia progenitors at explosion is only
broadly consistent with that predicted by Hachisu \& Kato (2001) for
systems with red-giant donors. Their models produce no SNe Ia with
final orbital periods ${\rm 2.5~d~\lesssim}~P_{\rm orb}~{\rm
\lesssim~60~d}$; most of the final parameter space for red-giant
donors has ${\rm 100~d~\lesssim}~P_{\rm orb}~{\rm
\lesssim~1000~d}$. Our favoured assumptions seem to indicate orbital
periods on the shorter side of 100 d, i.e. the kinematic signal seems
to be a little stronger than expected. A possible resolution could be
that our use of UCWDs is biased and preferentially selects systems
with shorter orbital periods because they produce lower-mass WDs,
which cool more rapidly. Or perhaps the progenitor population is
preferentially from the thick disc, and hence the remnants falsely
appear kinematically hotter than in our models. It may also be that
some non-LMWD single UCWDs from the thick disc or Galactic halo could
be contaminating these results. Alternatively, this may be an
indication that the models need some modification; for example the
systems containing red-giant donors might produce SNe Ia earlier than
expected, or perhaps a subset of systems from the supersoft channel
does not explode until their donor stars have evolved more than the
current models predict.

\subsection{The nature of UCWDs}

We encourage further work on the nature of UCWDs. It is important to
confirm that the apparently single UCWDs really are single. High
signal-to-noise searches for any spectral features that allow
radial-velocity measurements would be worthwhile.  GAIA astrometry is
a long-term hope for examining the single status of UCWDs, as well as
providing accurate distances to all these objects.

It is also important to understand the mass distribution of the UCWD
population. The inferred mass of the best-studied UCWD, LHS 3250,
($\rm 0.23~M_{\odot}$ [Bergeron \& Leggett, 2002]) makes it a clear
LMWD and hence a good candidate SN Ia remnant, but no other apparently
single UCWD has a well-constrained mass estimate that we are aware of.

\subsection{Outlook: a definitive signature?}

A sample of non-UCWD single LMWDs has distinct advantage over the
UCWDs in that the prescence of spectral lines allows for radial
velocity measurements (see, e.g., Maxted et al.\ 2000). The
significant difficulty for these objects is in selecting a large
sample: whilst UCWDs stand out from survey photometry, hotter LMWD do
not. Eisenstein et al.\ (2006) used the Sloan digital sky survey to
produce a catalogue of over 9000 white dwarfs, and identify 13 WDs
with masses $\rm < 0.3 ~{\rm M_{\odot}}$, of which 7 have masses $\rm
\leq 0.2 ~{\rm M_{\odot}}$.  These objects should be investigated for
signs of a companion NS, or for radial velocity variations.

With a substantial sample of single LMWDs, it would make sense to use
directional information about the space velocities of the objects
rather than just the magnitudes of the transverse velocities. We
expect that the use of such information will help distinguish between
a large initial scale-height, long orbital period population and a
thinner, shorter-period population. Our conclusions would be
considerably stronger if we were sure about the initial kinematics of
these stars; it is important to try and determine whether the remnants
come from the thick disc.

Once the single LMWD sample becomes large enough, it could potentially
be split into sub-samples with different WD masses. If our model is
correct, then less massive single LMWDs should be kinematically hotter
than more massive single LMWDs (see section
\ref{sec:OrbitalVelocities}).

Since we began this work, van Leeuwen et al.\ (2007) and Kilic et
al. (2007a) have searched for companions to LMWDs in the radio and
optical wavebands, respectively. The search by van Leeuwen et al.\ of
LMWDs for radio pulsations found none `down to flux densities of
0.6\,--\,0.8 mJy kpc$\rm{^{-2}}$', and concluded that `a given
low-mass helium-core white dwarf has a probability of $<$
0.18$\pm$0.05 of being in a binary with a radio pulsar'. For four WDs
with masses $\rm < 0.2 ~{\rm M_{\odot}}$, Kilic et al.\ found: `None
of these white dwarfs show excess emission from a binary companion,
and radial velocity searches will be necessary to constrain the nature
of the unseen companions.'  Our paper suggests that the
{\emph{assumption}} that there are unseen companions is not nessecary.

\subsection{Runaway hot subdwarfs}

Perhaps the most interesting possibility for the evolutionary state of
the donor in a single-degenerate SN Ia is that of a hot subdwarf (sdO
or sdB) star.  This would naturally allow short orbital periods ($\rm
\sim 1~h$) and also naturally produce extremely low-mass WDs, as
recently observed in the runaway WD LP 400-22 (Kawka et al.\ 2006),
which is inferred to have a mass of 0.17 $\rm ~{\rm M_{\odot}}$ and a
tangential velocity of 414 $\rm \pm$ 43 km $\rm {\rm s^{-1}}$.  This
would also provide a natural explanation for stars like the runaway
hot subdwarf US 708 (discovered by Hirsch et al.\ 2005). We feel that
a SN Ia origin for this object is more satisfying than a scenario
combining dynamical ejection from the supermassive black hole in the
Galactic centre with the simultaneous merger of two helium WDs (as
speculated by Hirsch et al.). The orbital velocity in this case may
well be augmented by a kick due to an impulse from the supernova shock
(e.g. Marietta et al.\ 2000).

The evolution of the WD--sdB binary system KPD 1930+2752 (see, e.g.,
Maxted, Marsh \& North 2000) has been investigated by Ergma, Fedorova
\& Yungelson (2001). They conclude that this system is likely to
eventually result in a merger of two WDs (see also Geier et
al. 2007). However we see no reason why similar systems could not
produce a SN Ia via a single-degenerate channel, hence producing such
objects as US 708 and then LP 400-22.

\section[]{Summary and Conclusions}

We have considered the formation of apparently single LMWDs in
general, concluding that the most natural scenario for the formation
of single LMWDs is that they are the remnants of donor stars in
single-denenerate SNe Ia. Indeed, lone LMWDs should be
{\emph{expected}} if some single-degenerate SNe Ia do occur with giant
donor stars, as inferred from the observations of Patat et al.\
(2007), notably if the donors lose a significant fraction of their
envelopes, as predicted for giant donors (Marietta et al.\ 2000). The
observations of Maxted et al.\ (2000), van Leeuwen et al.\ (2006) and
Kilic et al.\ (2006) are all in support of the existence of a
population of genuinely single LMWDs.

It seems difficult for the majority of apparently single UCWDs to
posess companions, and we have adopted them as a useful sample of
single LMWDs. We have integrated a population of SN Ia donor remnants
through a simple Galactic potential and compared the results of those
calculations to the known space velocities of apparently single UCWDs.
Our results are consistent with the single low-mass UCWDs having once
been red-giant donor stars at the time of a SN Ia explosion, as
predicted for single LMWDs.

A unified picture emerges in which the high-velocity WDs are remnants
of main-sequence donors in SNe Ia (as suggested first by Hansen,
2003), and a kinematically cooler population of single LMWDs were once
giant donors in long-period SN Ia progenitors: their longer orbital
periods led to a lower runaway velocity wheras their tenuous envelopes
were stripped more easily by the supernova ejecta to produce LMWDs.

Furthermore, it seems plausible that runaway LMWDs such as LP 400-22
and runaway hot subdwarf stars such as US 708 originate from donor
stars in short-period ($\rm \sim 1~h$) SN Ia systems. We will explore
this idea in more detail in a future paper.

\section*{Acknowledgements}

We thank Uli Heber for very interesting conversations and for bringing
the issue of runaway hot subdwarfs to our attention. Questions from
Marten van Kerkwijk helped improve the clarity of our arguments, and
we thank an anonymous referee for their useful comments. Discussions
with the stellar group at Oxford were also useful.

SJ has been supported by PPARC grant PPA/G/S/2003/00056 \& Global Jet
Watch, and CW by a PPARC Advanced Fellowship. ZH visited Oxford
thanks, in part, to a Royal Society UK-China Joint Project Grant
(Ph.P. and Z.H.). This work was partly supported by the National
Science Foundation of China under Grant Nos. 10521001, 10433030 and
2007CB815406 (Z.H.) and a European Research \& Training Network on
Type Ia Supernovae (HPRN-CT-20002-00303).

\label{lastpage}
\end{document}